\documentclass[sigconf]{acmart}

\usepackage[utf8]{inputenc} 
\usepackage{graphicx} 
\usepackage{booktabs} 
\usepackage{array} 
\usepackage{paralist} 
\usepackage{verbatim} 
\usepackage{algorithmicx}
\usepackage{setspace}
\usepackage[ruled,vlined,linesnumbered]{algorithm2e}
\usepackage{epstopdf}
\usepackage{longtable}
\usepackage{subfigure}
\usepackage{color}
\usepackage{url}
\usepackage{epstopdf}
\usepackage{amsmath}
\usepackage{amsmath,amssymb,latexsym,graphics,epsfig,psfrag,float}
\usepackage{fancyvrb, comment}

\setcopyright{rightsretained}

\acmConference[KDD'18]{ACM Conference on Knowledge Discovery and Data Mining}{August 2018}{London, UK}
\acmYear{2018}
\copyrightyear{2018}

\acmArticle{4}
\acmPrice{15.00}

\begin{document}
\title{Information of Epileptic Mechanism and its Systemic Change-points in a Zebrafish's\\ Brain-wide Calcium Imaging Video Data}

\author{Jingyi Zheng}
\affiliation{%
  \institution{University of California, Davis}
  \city{Davis}
  \state{California}
  \postcode{95616}
}
\email{jgzheng@ucdavis.edu}

\author{Fushing Hsieh}
\authornote{Correspondence}
\affiliation{%
  \institution{University of California, Davis}
  \city{Davis}
  \state{California}
  \postcode{95616}
}
\email{fhsieh@ucdavis.edu}


\begin{abstract}

The epileptic mechanism is postulated as that an animal's neurons gradually diminish their inhibition function coupled with enhanced excitation when an epileptic event is approaching. Calcium imaging technique is designed to directly record brain-wide neurons activity in order to discover the underlying epileptic mechanism. In this paper, using
one brain-wide calcium imaging video of Zebrafish, we compute dynamic pattern information of the epileptic mechanism, and devise three graphical displays to show the visible functional aspect of epileptic mechanism over five inter-ictal periods. The foundation of our data-driven computations for such dynamic patterns relies on one universal phenomenon discovered across 696 informative pixels. This universality is that each pixel's progressive 5-percentile process oscillates in an irregular fashion at first, but, after the middle point of inter-ictal period, the oscillation is replaced by a steady increasing trend. Such dynamic patterns are collectively transformed into a visible systemic change-point as an early warning signal (EWS) of an incoming epileptic event. We conclude through the graphic displays that pattern information extracted from the calcium imaging video realistically reveals the Zebrafish's authentic epileptic mechanism.

\end{abstract}

%
%



\maketitle

\section{Introduction}

Epilepsy is a neurological disorder in the brain characterized by recurrent, unprovoked seizures. As the most common type of epilepsy, idiopathic epilepsy has no identifiable cause. 
Nowadays, idiopathic epilepsy is still not preventable. To better prevent epilepsy, it is essential to understand the underlying mechanism that the central nervous system (CNS) change leads to epileptic seizure. However, this seemingly natural research direction has not prevailed in literature.

Early works on epileptic seizure prediction are mainly based on Electroencephalography (EEG) data. For example, Rogowski {\em et al.} \cite{rogowski} applied autoregressive model on 8-channel EEG signal recorded from 12 epileptic patients and claimed the trajectory of some prediction coefficients during pre-seizure period could predict epileptic seizures by several seconds. Salant {\em et al.} \cite{salant} detected EEG changes using multivariate spectral estimation on 2-channel EEG signal and used coherence and pole trajectories to predict epileptic seizures. Some other studies tried to predict epileptic seizure using spike occurrence rates in the EEG \cite{gotman, koffler, katz, martinerie}. As an example, Le Van Quyen \cite{levan, levan2} later developed dynamical similarity index and found decrease in dynamical similarity, compared to a pre-ictal reference constant, before seizures.


While these studies only focus on pre-ictal period, not inter-ictal period,  Navarro \cite{navarro} showed that similarity measure drops more frequently in pre-ictal period than inter-ictal period. Aschenbrenner {\em et al.} \cite{Aschenbrenner-Scheibe} also found that there is no significant difference in dimension drop between pre-ictal and inter-ictal period. On the other hand, many research work explored the localized epileptic channels \cite{ElgerandLehnertz, martinerie, D'Alessandro, mormann, Esteller} and showed that the remote channels could carry relevant information about epileptic seizures. Using EEG recordings from a small group of patients with mesial temporal lobe epilepsy, people could predict seizure onset around 20 minutes in advance, but recording EEG signal requires continuous and invasive intracranial recordings. EEG recordings are only from limited number of channels, the underlying neuron network is hard to capture. Hence, it is curial to systemically study the epileptic mechanism to identify and predict the epileptic seizure.
%


Generally, the epileptic mechanism is heuristically postulated as that a large number of neurons experience decreasing inhibitions and increasing excitations simultaneously on the road to an epileptic event. However, there is no existing visible evidence on such mechanistic characteristics reported in literature yet. Meanwhile, if epileptic event is a collective behavior of neurons in CNS, epileptic mechanism should be better manifested through large number of neurons, rather than a small number of channels. Further aforementioned analysis methodologies are primarily based on modeling the EEG time series, instead of focusing on extracting authentic information pertaining to the epileptic mechanism. To address those problems, one recent advance in brain-wide calcium imaging has made possible to graphically reveal collective patterns of epileptic mechanism. Particularly, the calcium imaging technologies are equipped with higher sampling rate than functional magnetic resonance imaging (fMRI), and produce a more precise brain-wide imaging than limited number of EEG electrodes.

In this paper, we study the epileptic mechanism using a zebrafish's brain-wide calcium imaging video data. The larval fish is immersed in a drug that induced epileptic event, mimicking epilepsy. The high-speed confocal microscopy is used to image larval zebrafish. In contrast with EEG recording from limited number of channels, the calcium imaging video data is capable of monitoring brain-wide network activity using non-invasive monitor.
We first reduce the video data dimension using K-variance method. Upon each selected pixel, we devise a progressive quantile-statistics to show individual pixel's mechanistic patterns while an epileptic event is approaching. Then we accordingly encode the calcium fluorescence intensity time series into a digital series as a way of re-normalizing all time series data. This is one essential step for graphically revealing epileptic mechanism on the overall global scale.

We then aggregate digital encoded recording from each pixel into a matrix. This matrix representation clearly and naturally demonstrates the evolving systemic patterns from the beginning to the end of inter-ictal period. We then compute all local linear trends upon all pixels' progressive quantile-statistics along the temporal axis within each inter-ictal period via a moving window. Finally, we develop an algorithm to collectively summarize all local trends into a time series that could successfully reveals a critical systemic change-point, and then predict the onset of next epileptic event.

The remainder of the paper is organized as follows: We introduce the data set and data processing in Section~\ref{MM}. We present our methodologies in Section~\ref{analysis}. In Section~\ref{Results}, we show results to validate our findings. We conclude this paper in Section~\ref{con}.

\section{Data Processing}\label{MM}

In this section, we first introduce the video data, then present K-variance method to reduce the dimension of the data, and show the two key components: progressive quantile statistic to reveal pixel-wise epileptic patterns and digital encoding to standardize calcium fluorescence intensities.

\subsection{Data}

The epileptic mechanism is heuristically postulated as that a large number of neurons experience decreasing inhibitions and increasing excitations simultaneously on the road to an epileptic event. Biologically, neurons convey information using spikes \cite{ganmor2016direct}. To measure neural activities, medical imaging technologies are developed to create visual representations of the interior of a body for clinical analysis and medical intervention \cite{cook2007medical, chakraborty2011new}. Particularly, two-photon imaging of calcium indicators, which is designed to show the calcium status of an isolated cell, are used for monitoring the neuronal activity in hundreds of distinct neurons \cite{denk1990two, stosiek2003vivo}.

\begin{figure}
\centering
 \includegraphics[scale=0.31]{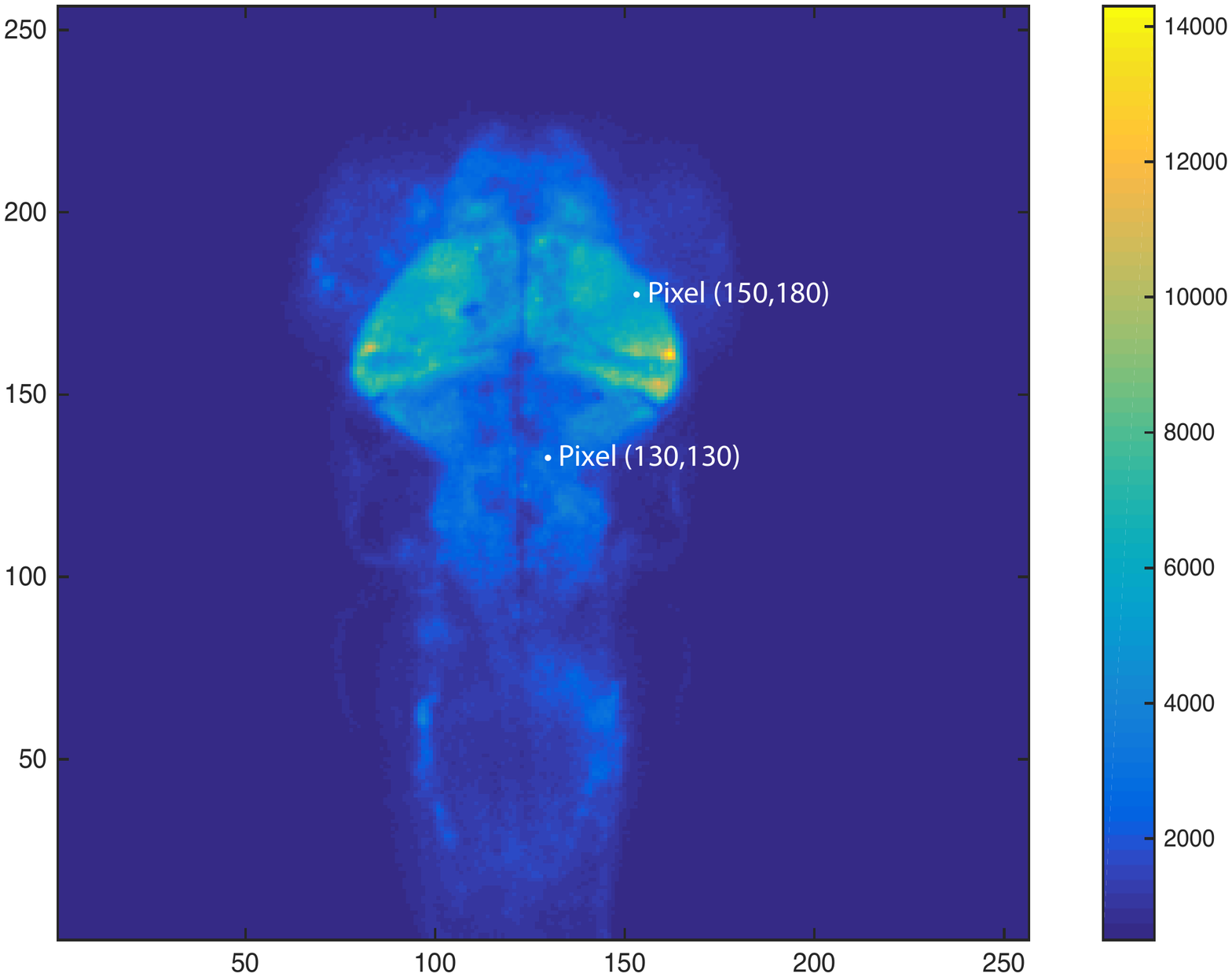}
   \caption{Zebrafish Brain Image Data}
 \label{fig:data}
\end{figure}

To study the epileptic mechanism, we use a zebrafish's brain-wide calcium imaging video data from Epilepsy Research Laboratory at University of California, San Francisco. To collect the imaging data, a well characterized seizure induction protocol, Pentylenetetrazol (PTZ), is used in the experiment. PTZ is a common convulsant agent to induce epileptic seizures. The larval fish is immersed in PTZ to mimic epilepsy. The high-speed confocal microscopy is used to image larval zebrafish, and electrical discharge is recorded by calcium fluorescence intensity (CFI). As an example, one frame of the video data is shown in Figure \ref{fig:data}. In contrast with EEG recording from limited number of channels, this data is capable of monitoring brain-wide network activity using non-invasive monitor.

The frame rate of the video data is $6.7 Hz$, and there are $2000$ frames in total. The resolution for each frame is $256*256$. Thus, the video consists of $2^{16}$ pixel-specific CFI time series.

\subsection{Dimension Reduction}

\begin{figure}
  \centering
  \subfigure[Calcium Fluorescence Intensity Series -- Body]{\includegraphics[scale=0.31]{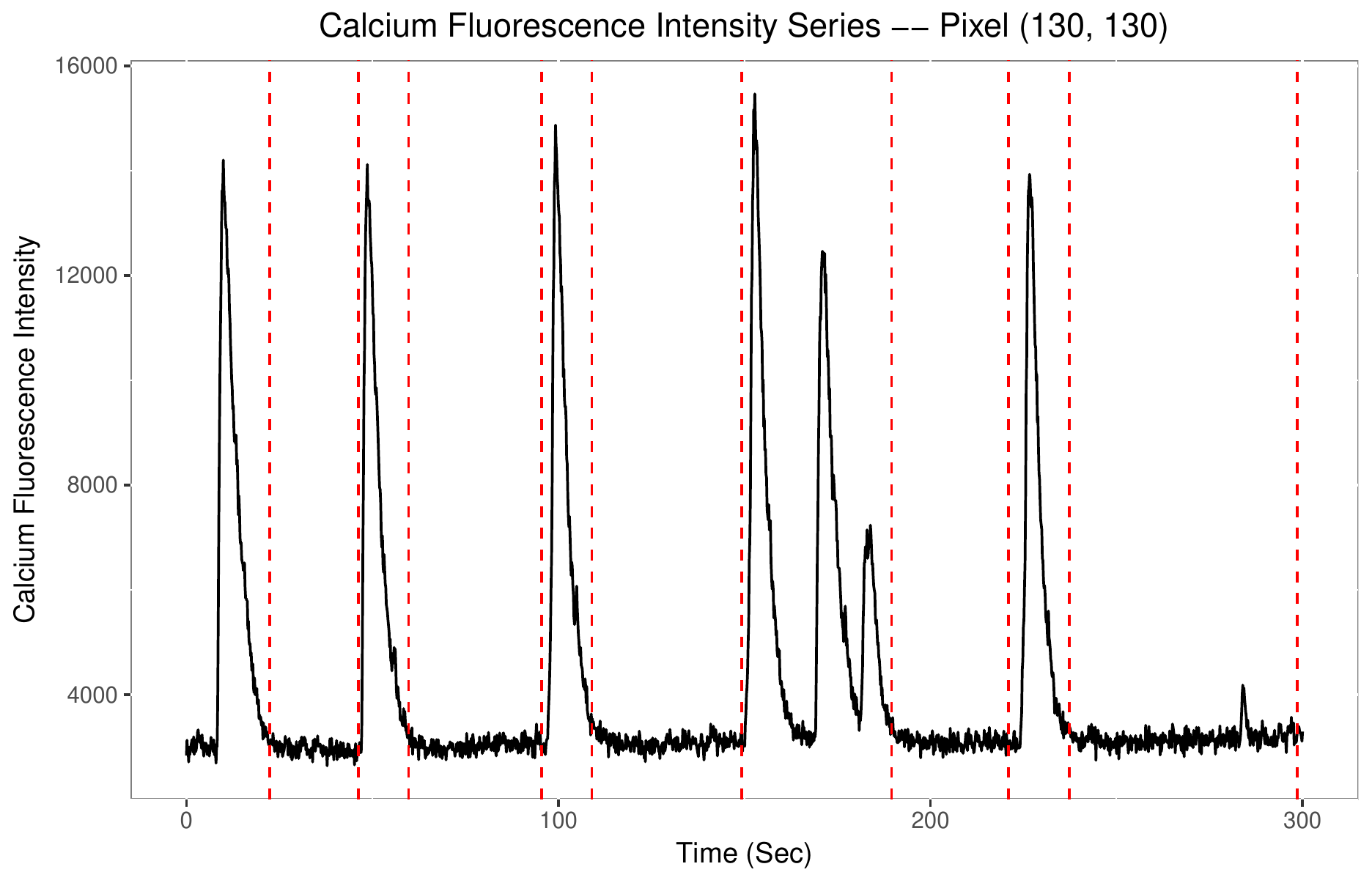}}
  \subfigure[Calcium Fluorescence Intensity Series -- CNS]{\includegraphics[scale=0.31]{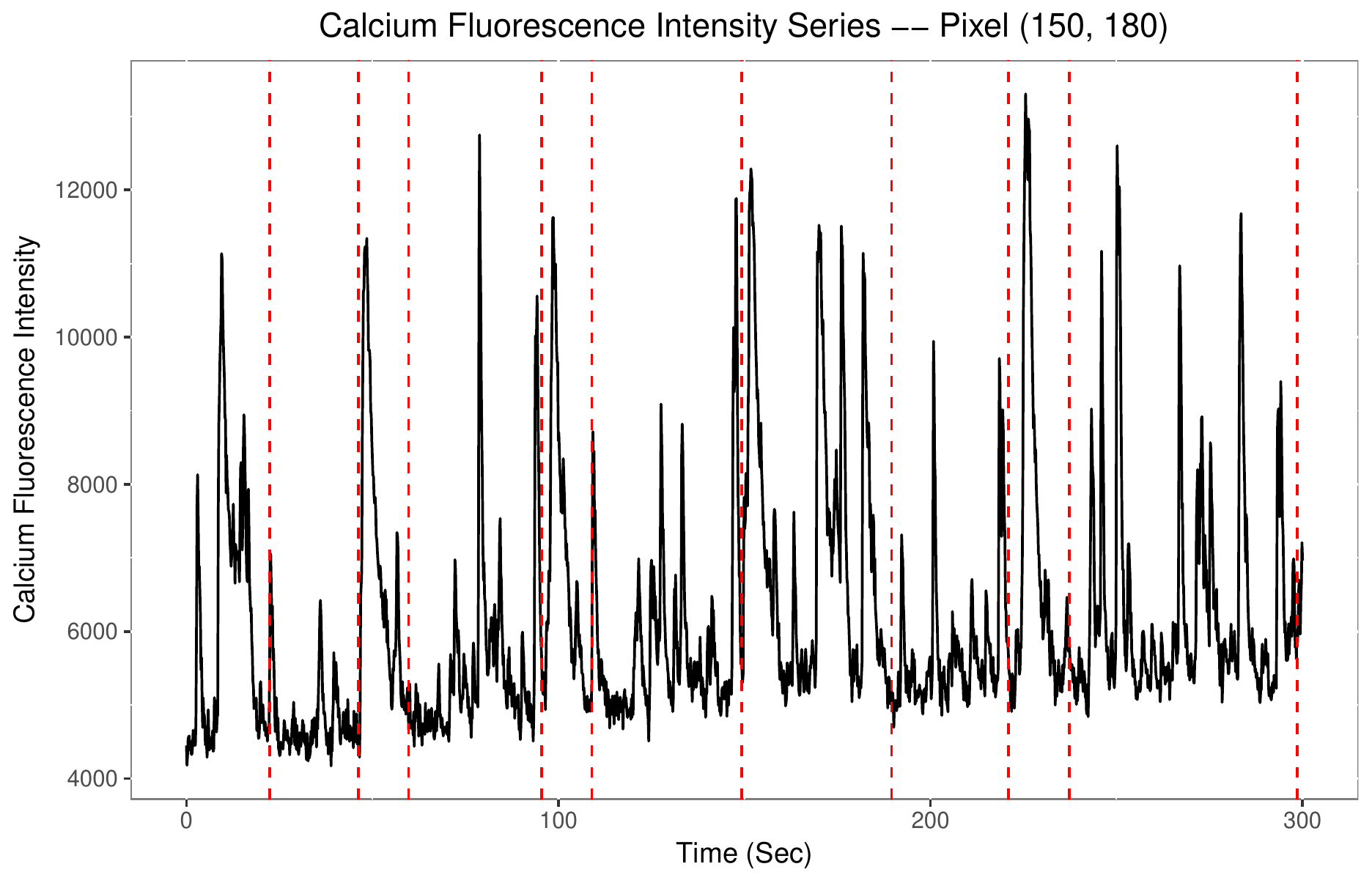}}
  \caption{Calcium Fluorescence Intensity Imagine Data}
    \label{fig:Fig1}
\end{figure}
Our Zebrafish's brain-wide calcium imaging video data contains total 2000 frames of $256*256$ images. Among $256*256$ pixel-specific time series, we observe two main types of pixels based on their CFI trajectories. Figure \ref{fig:Fig1} shows two distinct CFI trajectories. Panel (a) of Figure \ref{fig:Fig1} shows a CFI series of one pixel taken from the body part of zebrafish, and panel (b) shows a CFI trajectory of one pixel taken from the brain of zebrafish. The location of those two pixels are marked in Figure \ref{fig:data}. The CFI trajectory in panel (a) has nearly no fluctuation between spikes, and only spikes can be clearly seen. In other word, CFI series like panel (a) only reflects epileptic seizures. On the other hand, the CFI in panel (b) not only reflects epileptic events, but also gives out high fluctuation across each inter-ictal segment.

Two types of CFI series share one common feature: spikes that indicate epileptic seizures. The difference is the fluctuation during inter-ictal segments, which is essential to reveal the connectivity among neurons in CNS. The fluctuation between epileptic seizures can be used to discover interacting relations with other pixels and the dynamic mechanism leading to epileptic seizure. For the purpose of prediction, we define pixels with CFI series like panel (b) as informative pixels and pixels with CFI series like panel (a) as non-informative pixels.

CFI series of non-informative pixels could be used as a basis for partitioning and separating inter-ictal segments from whole time series. We randomly pick up ten pixels from zebrafish body. Then we build empirical distribution of its CFI for each pixel, and choose $70^{th}$ percentile as a cut-off point. Time points with intensity greater than $70^{th}$ percentile are labeled as epileptic seizure. Then we decide the overall epileptic seizure time based on consensus voting. Totally we have five inter-ictal segments since we combine the three consecutive spikes as the $4^{th}$ epileptic seizure and treat the last segment as inter-ictal segment.

\begin{figure}[!tbp]
  \centering
    \includegraphics[width=0.4\textwidth]{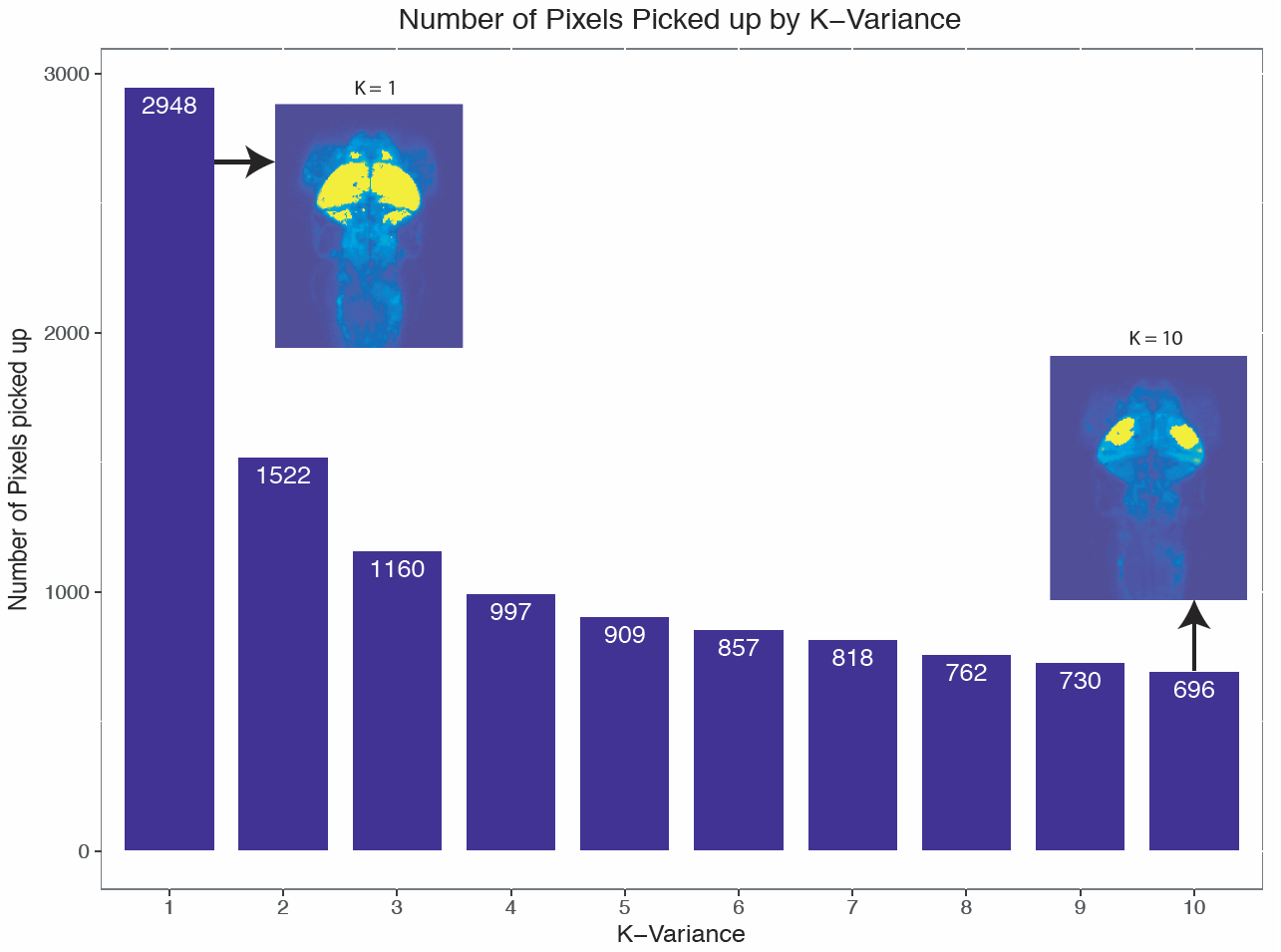}
    \caption{K-variance  Algorith}
    \label{fig:Fig2}
 \end{figure}

\begin{figure}[!tbp]
  \centering
    \includegraphics[width=0.33\textwidth]{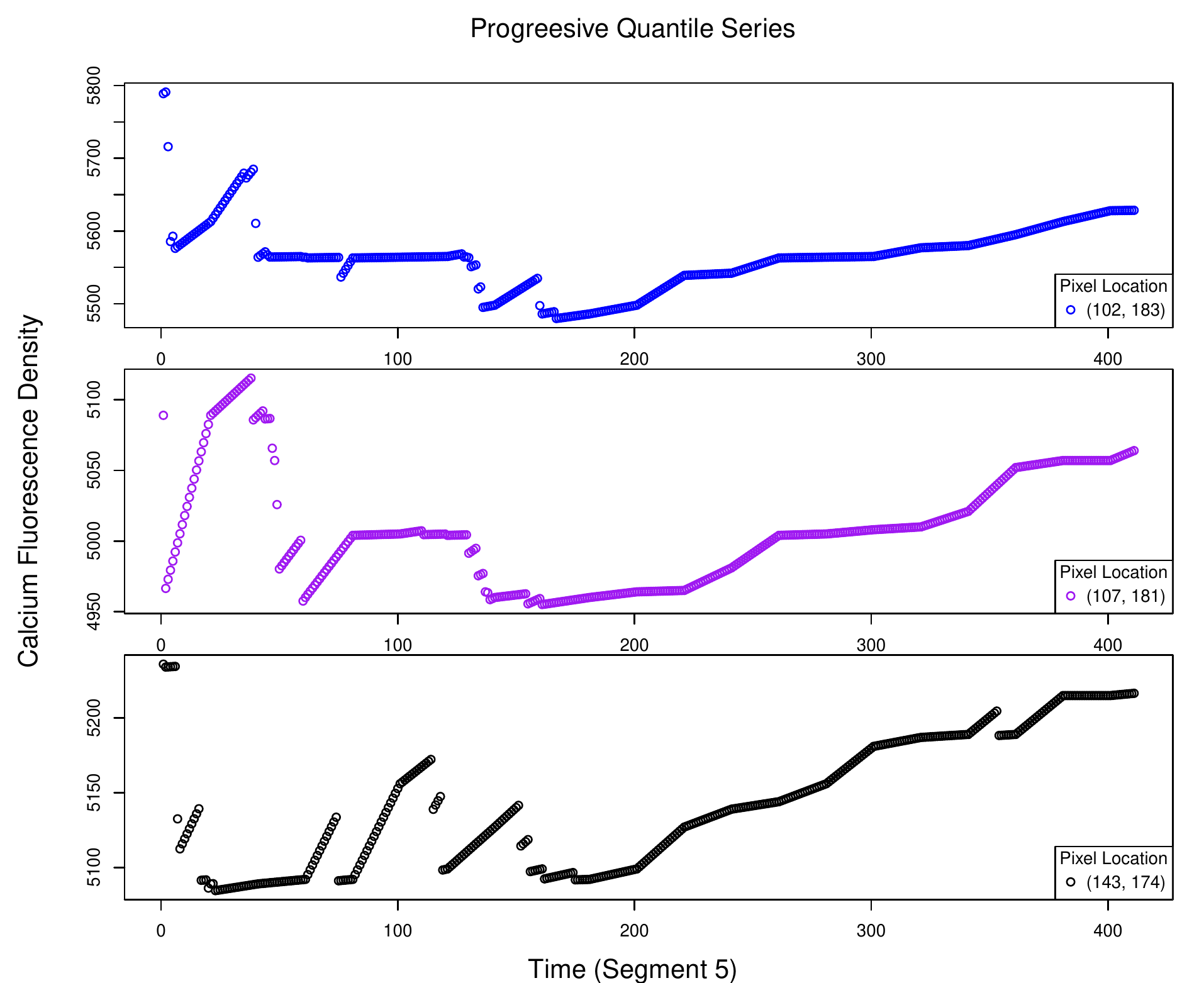}
    \caption{Progressive Quantile Series}
    \label{fig:quantile}
 \end{figure}

To pick up informative pixels, we develop a K-variance method. 15 pixels from zebrafish body are randomly picked up and partitioned into five inter-ictal segments. Intensity variance for each pixel during five inter-ictal segments $[v_{i1}, v_{i2}, v_{i3}, v_{i4}, v_{i5}]$ is computed.\\
$$[v_{i1}, v_{i2}, v_{i3}, v_{i4}, v_{i5}]$$  for $i^{th}$ pixel, $ i = 1,2,3,\cdots,15 $
$$V_j = mean_{i}(v_{ij}), i = 1,2, \cdots,15, j = 1,2,\cdots,5$$

For inter-ictal segment $j$, the mean of 15 variances $v_{ij}, i=1,\cdots,15$ is taken as the baseline $V_j$. Then we go through the CFI series of all $256*256$ pixels, and partition them into $5$ inter-ictal segments, calculate intensity variance, and compare with the baseline for each segment. If variances for all segments are greater than baselines, the pixel is labeled as informative pixel. As shown in Figure \ref{fig:Fig2}, we have collected the informative pixels after scanning $2^{16}$ CFI series (marked with $k=1$). The dimension of data is reduced to 2948 CFI, and the target area is narrowed down to the brain of zebrafish instead of the whole body.

To get a even smaller target area, we increase the lower bound $V_j$ to $k*V_j, j=1,\cdots,5$. With greater $k$, less pixels are collected as informative, and target area is narrowed down to the optic tectum.  Figure \ref{fig:Fig2} shows the number of pixels that are picked up using K-variance algorithm. When $k$ is increased to $10$, we have 696 informative pixels, and our target areas mainly focus on spinal trigeminal (SpV) neurons and Neuropil.

\subsection{Progressive Quantile Statistics}\label{quantile}
Inhibition and excitation functionality of neurons are represented by CFI values in our video data. High intensity value indicates excitation while low value shows inhibition. As one pixel in optic tectum possibly involves more than one neuron, the inhibition and excitation that we discuss in this paper are meant to be surrogates of the biological ones.
When neuron system experiences intense inhibition, less neurons are active around one pixel at time $t$, which result in its CFI value is more likely to be low, or even in the trough overall. Therefore, we use a trough or a relatively low value of CFI as a surrogate for a pixel experiencing inhibition at time $t$, while a peak or a high value for excitation.

To discover the decreasing inhibition and increasing excitation mechanics, we propose to use a progressive quantile statistics. For ${i^{th}}$ pixel, we have a CFI series ${S_i(t)}, {0<t<T}$, recorded from $t=0$ to a time point $t$, where ${t = 0}$ is the beginning of inter-ictal segment, and ${T}$ is the end of inter-ictal segment, which is also the onset of next epileptic seizure. Let ${D_i(t)}$ be the cumulative empirical distribution of ${S_i(t)}$, and we use ${D_i^{-1}(0.05|t)}$ to denote empirical ${5\%}$ quantile of ${D_i(t)}$ at time t.

Figure \ref{fig:quantile} shows a typical trajectory of ${D_i^{-1}(0.05|t)}$ in our epilepsy setting. The initial bumpy period actually occurs in both epilepsy and control setting. On one hand, this is because the time is too short for ${D_i^{-1}(0.05|t)}$ to be stable around the beginning of inter-ictal segments. On the other hand, ${D_i^{-1}(0.05|t)}$ being dragged down during bumpy period is due to functioning inhibition, which produces low CFI value.

After the bumpy period, if inhibition and excitation function of neuron system works concordantly, as in the control setting, we would expect to see that the progressive ${D_i^{-1}(0.05|t)}$ converges to a constant value. This is indeed confirmed in control fish settings, as reported in the companion biological report.
If the system gradually loses inhibition and gains only excitation, then the cumulative ${D_i(t)}$ would has more high value coming in but no low value, which would result in an increasing ${D_i^{-1}(0.05|t)}$. This is actually the case of epilepsy setting as shown in Figure \ref{fig:quantile}.

The progressive quantile statistic ${D_i^{-1}(0.05|t)}$ trajectory is coherent with the mechanistic pattern leading to an epileptic event during inter-ictal segments. Based on this pattern formation, we could conclude that the 696 pixels are all informative for this zebrafish's epileptic mechanism.

\subsection{Digital Encoding} 

 \begin{figure}[htp]
  \centering
  \subfigure[]{\includegraphics[width=0.35\textwidth]{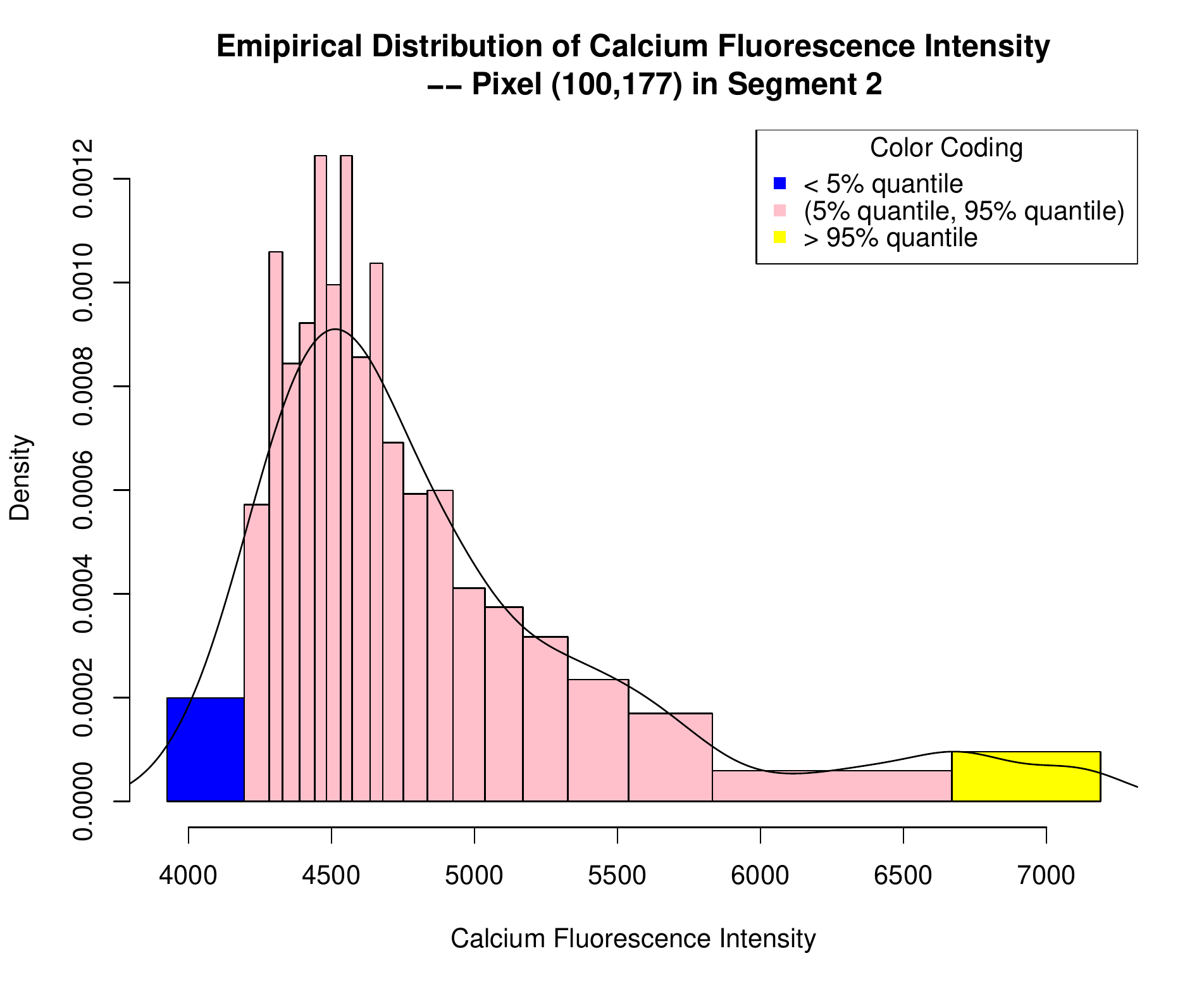}} \quad
  \subfigure[]{\includegraphics[width=0.43\textwidth]{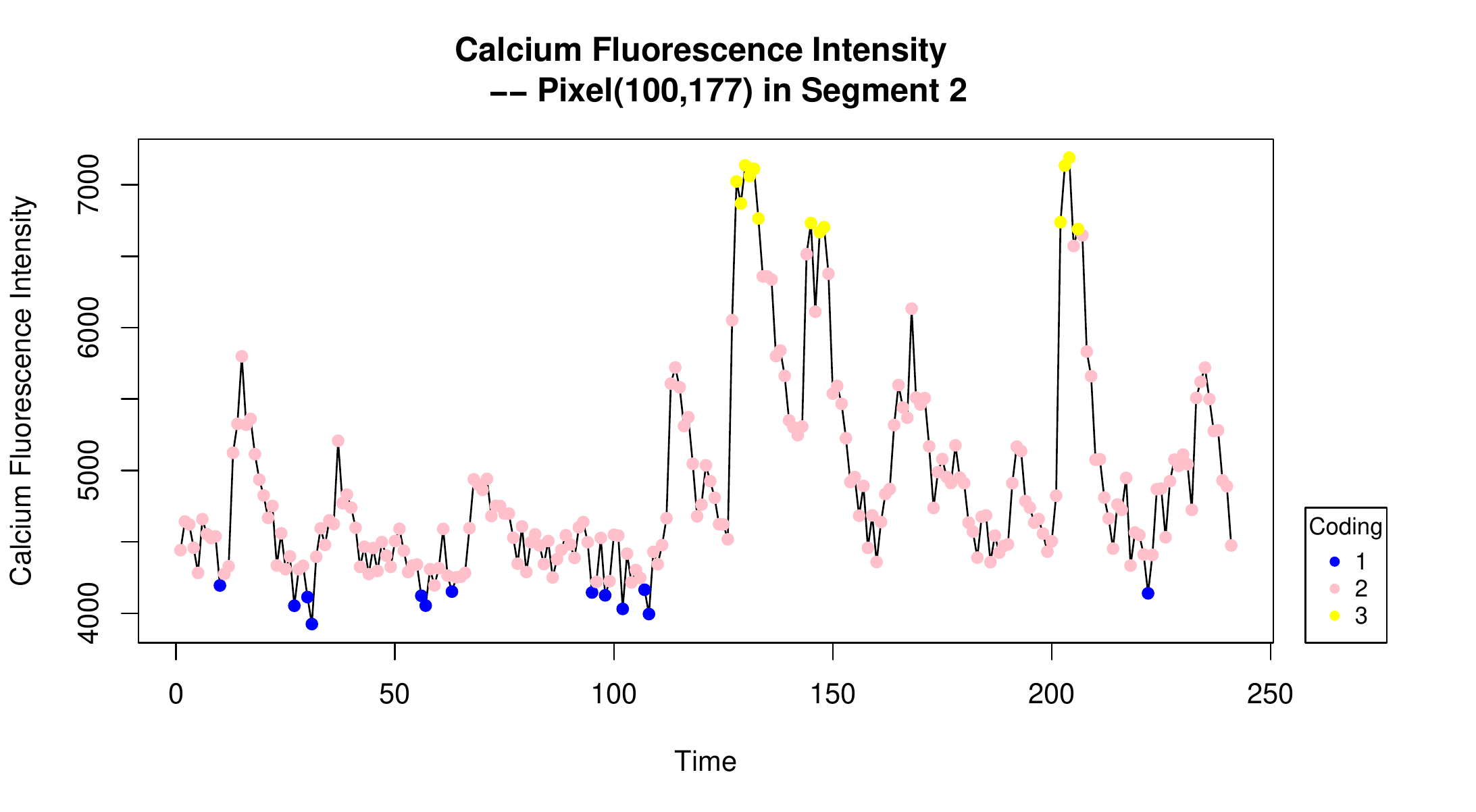}}
  \subfigure[]{\includegraphics[width=0.45\textwidth]{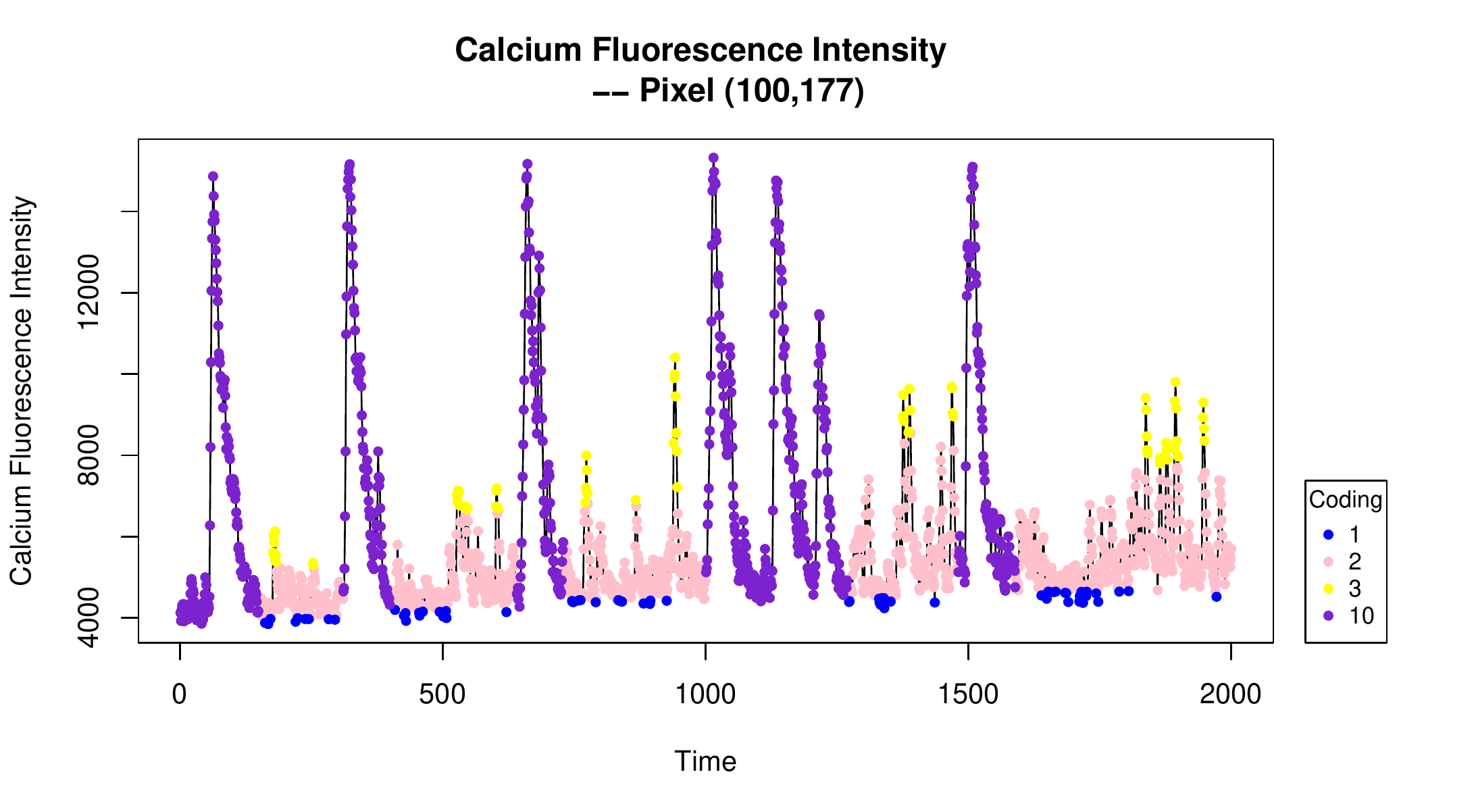}}
  \caption{Digital Coding Algorithm}
    \label{fig:Fig3}
\end{figure}

Among $696$ pixels, the range and scale of CFI series are really different. To standardize CFI series, we apply a digital encoding scheme on CFI series as following. For each pixel, we compute empirical distribution of the intensity series, partition the histogram into 3 sub-regions, and encode 3 sub-regions in a monotonic fashion as shown in Figure \ref{fig:Fig3} (a). We choose to encode raw CFI series into 3-digit coding $\{1, 2, 3\}$. The intensities that are less than its lower $5\%$ quantile in CFI series is coded as $1$ in our digital coding, intensities that are greater than its upper $5\%$ quantile, that is $95\%$ percentile, is coded as $3$, and the rest of intensities falling between lower and upper $5\%$ quantiles are coded as $2$.

Our triplet-encoding not only standardize all CFI series, but also retain information of inhibition and excitation mechanics within each pixel-wise CFI series. As shown in Figure \ref{fig:Fig3} panel (b), the triplet-encoding could capture peak and trough, which represent excitation and inhibition. Thus, each pixel's CFI series is transformed into a triplet-encoding sequence for five inter-ictal periods.

\begin{itemize}

  \item less than $5\%  \rightarrow 1 \rightarrow$ trough (color-coded blue)
  \item between $5 \%$ and $95 \% \rightarrow 2$ (color-coded pink)
  \item greater than $95\%  \rightarrow 3 \rightarrow $ peak (color-coded yellow)

\end{itemize}

Figure \ref{fig:Fig3} panel (c) shows five triplet-encoding series connected with epileptic seizure being coded as 10. We observe that majority of blue dots appear to be around the beginning of inter-ictal segments, while yellow dots are gathered around the end of segments when next epileptic seizure is about to happen. This observation explains the epileptic mechanics from one pixel perspective. It shows that decreasing inhibition and increasing in excitation actually occur when an epileptic event is approaching.

\section{Methods} \label{analysis}

In this section, we first aggregate the digital encoding series of $696$ pixels to discover the underlying epileptic mechanism, then detect the systemic change-point based on the changing pattern in CNS during inter-ictal period. Based on the systemic change-points, we develop an algorithm to predict the next epileptic seizure event.

\subsection{Digital Encoding Series Aggregation} 

\begin{figure}[htp]
  \centering
  \subfigure[Segment 1]{\includegraphics[scale=0.35]{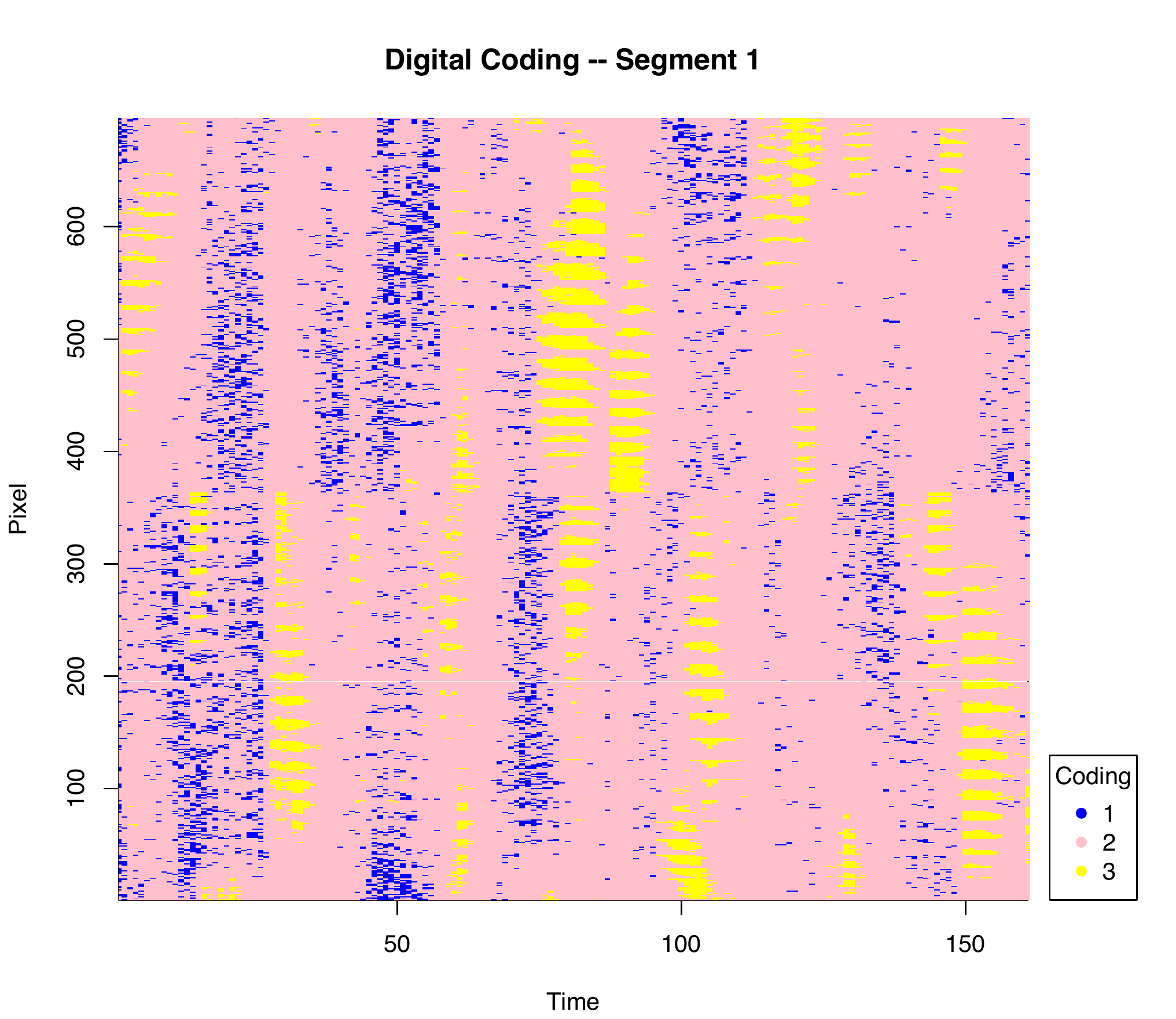}}
  \subfigure[Segment 2]{\includegraphics[scale=0.22]{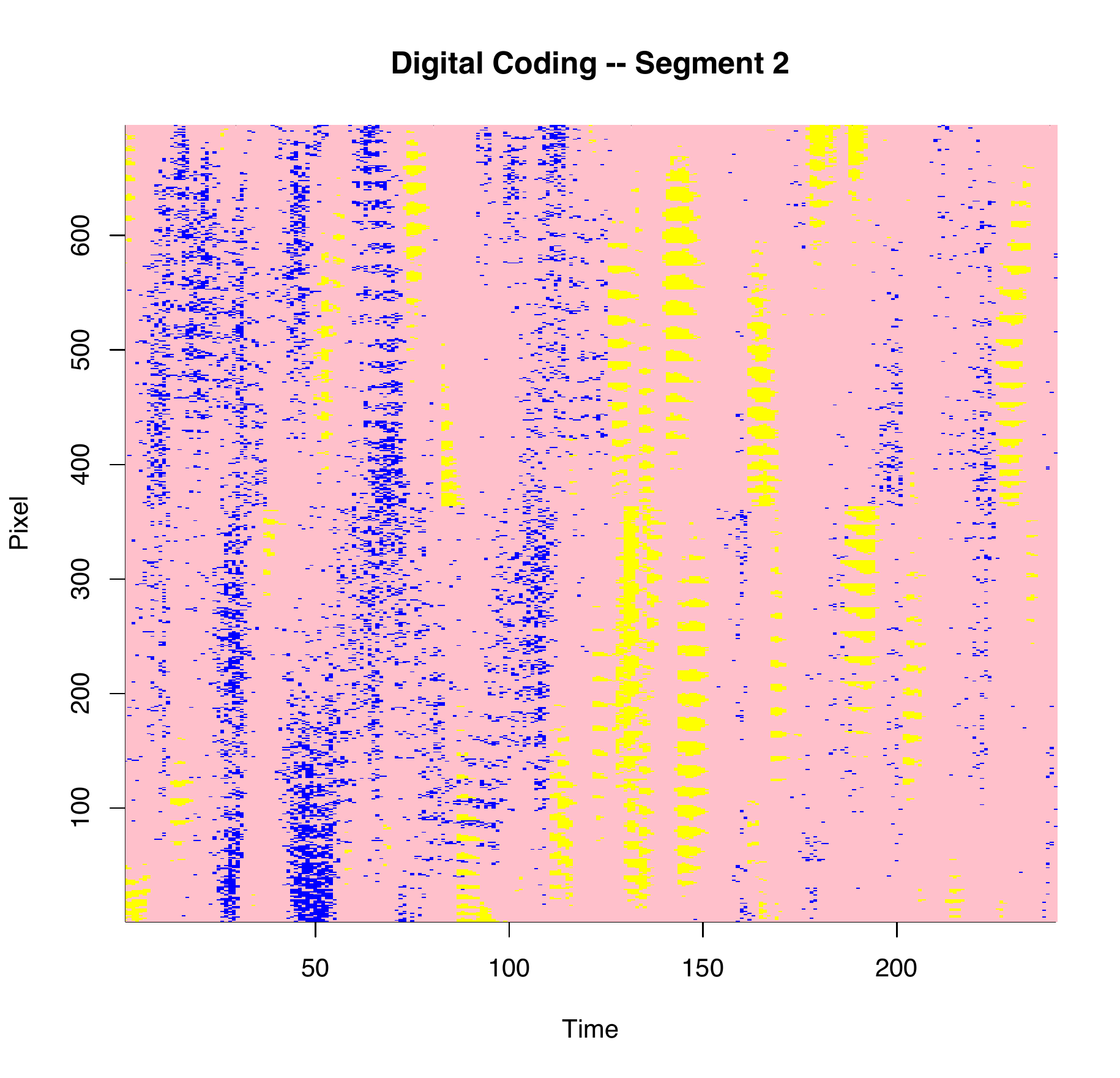}}
  \subfigure[Segment 3]{\includegraphics[scale=0.22]{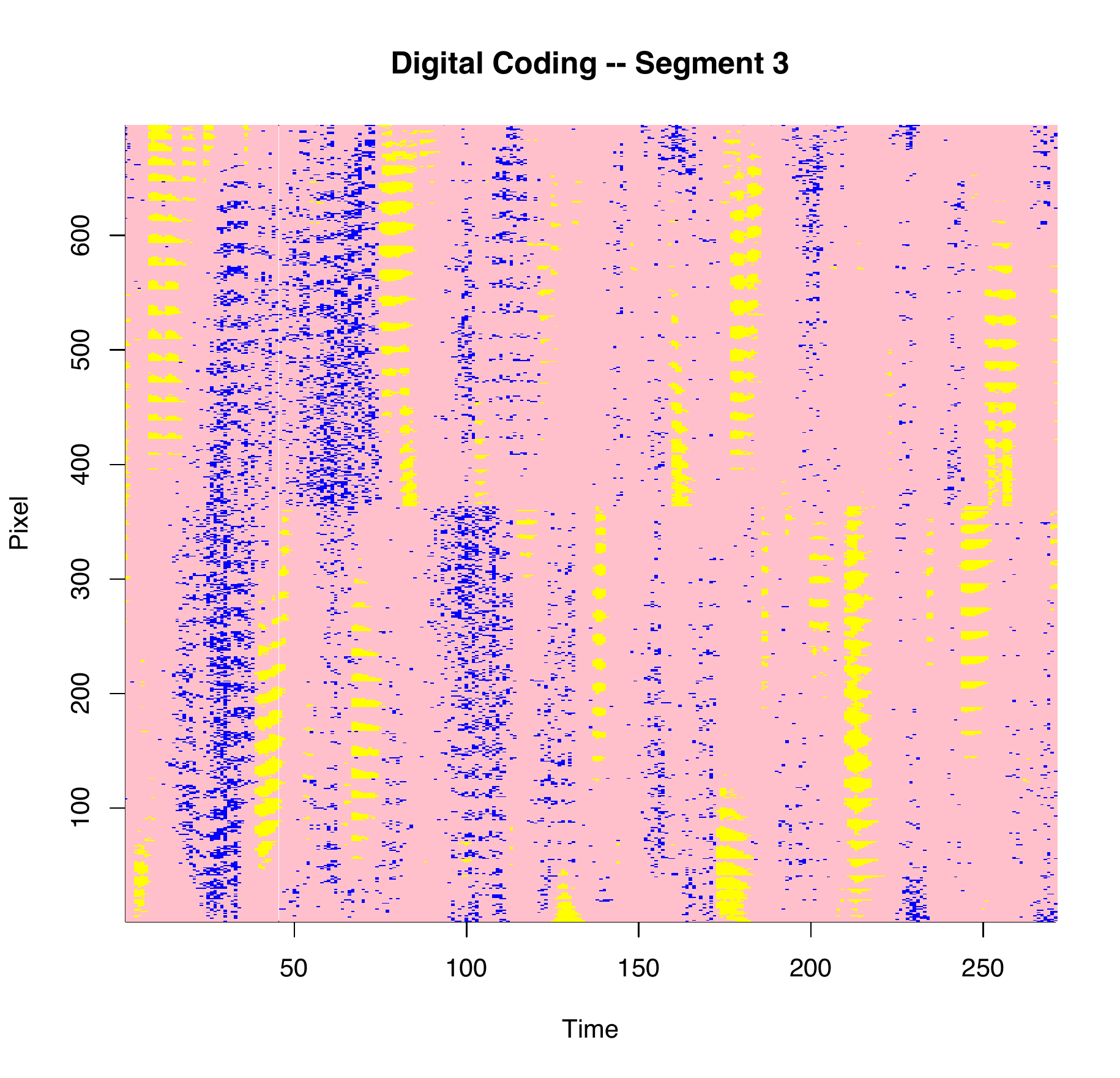}}
  \subfigure[Segment 4]{\includegraphics[scale=0.22]{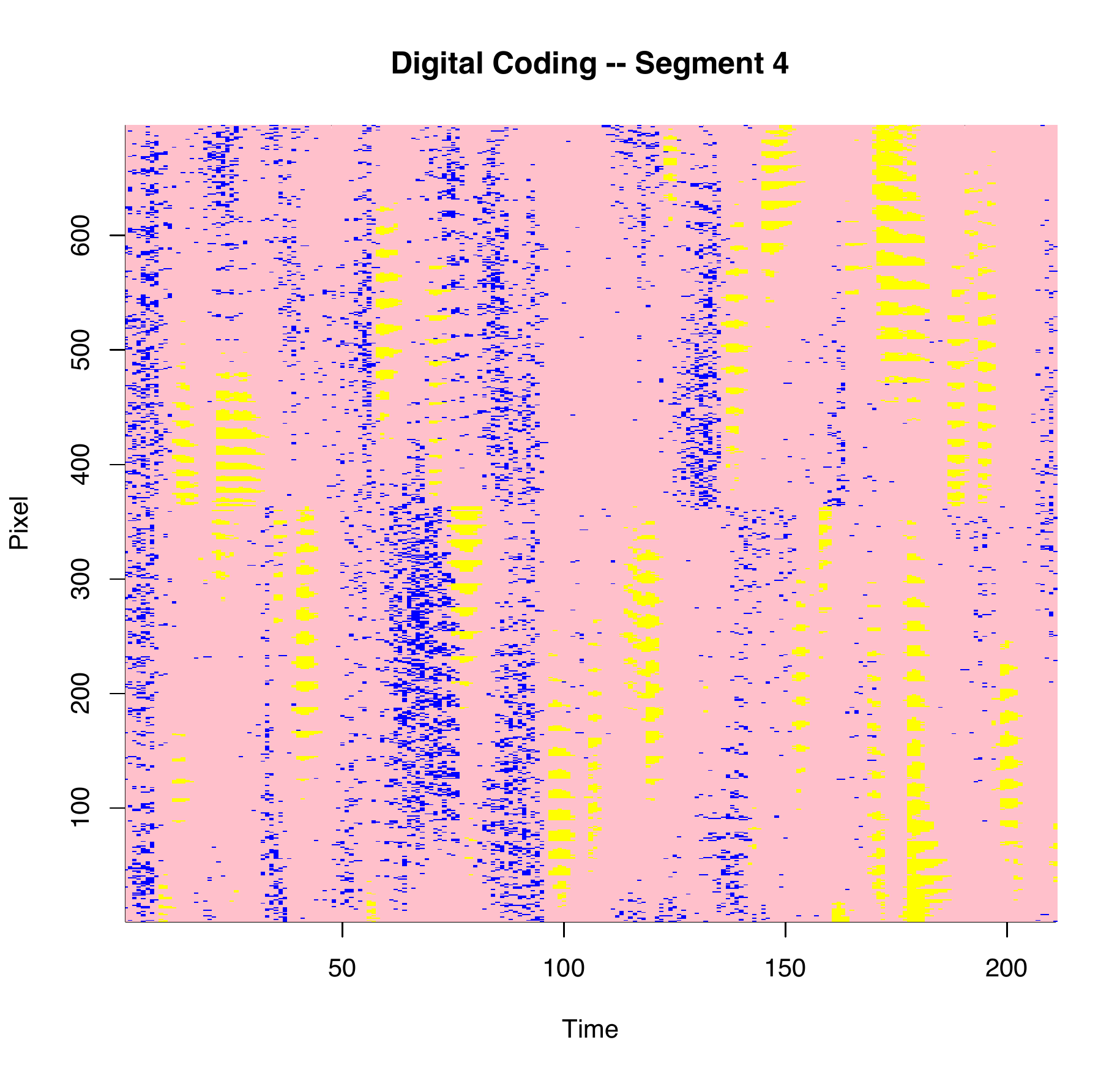}}
  \subfigure[Segment 5]{\includegraphics[scale=0.22]{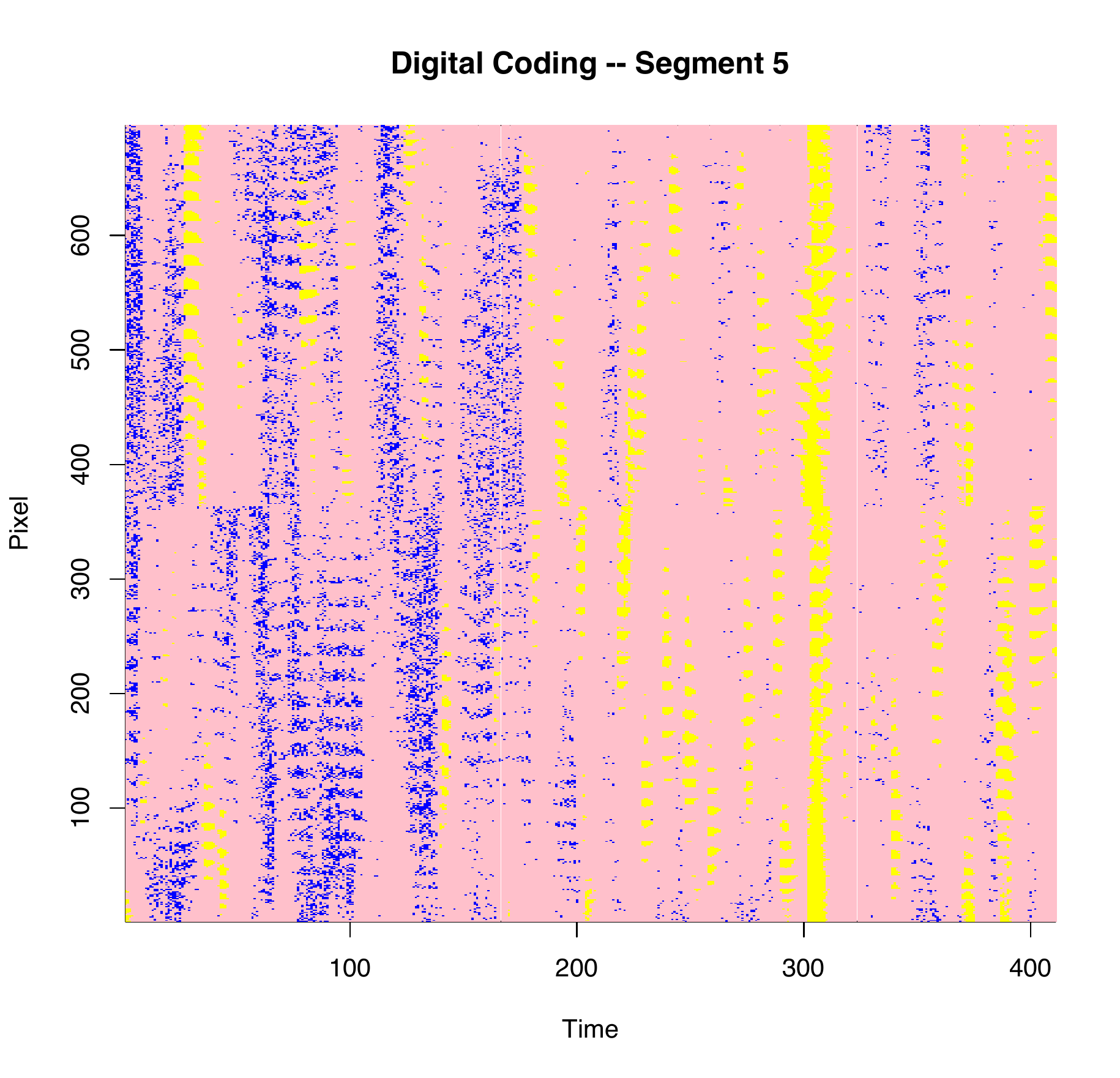}}
  \caption{Heatmaps of five inter-ictal periods: Graphic displays of systemic wax-and-wane patterns of troughs and peaks}
   \label{fig:heatmap}
\end{figure}

To explore the common pattern that all informative pixels share, we aggregate $696$ digital encoded sequences into a heatmap. Specifically, we index the pixels, and each row in the heatmap represents the digital encoded series for one pixel. The streaming imaging data shown in Figure~\ref{fig:data} is summarized into a matrix. The vertical axis is the pixel index and the horizontal axis shows the CFI change in time domain. As shown in Figure~\ref{fig:heatmap}, we conduct the aggregation for five inter-ictal segments. From the heatmaps of different segments, we have several observations:


\begin{itemize}
  \item There are clearly two different parts in all five heatmaps, upper part and lower part. According to the pixel arrangement, the lower part are from the left half of tectum, and the upper part are from the right half of tectum. With respect to temporal axis, both parts of tectum share similar patterns of blue and yellow dots. With respect to spatial axis, the neuron activities are somehow staggered between left and right part of tectum.
  \item The aggregation pattern of yellow dots implies synchrony among nearby neurons.
  \item The blue-dot-clouds are more intense during the beginning half of inter-ictal segment, and gradually disappear when next epileptic seizure is coming.
\end{itemize}
The third systemic pattern is proved very informative in next section.

\subsection{Systemic Change-point Detection} 
\begin{figure}[!tbp]
  \centering
    \includegraphics[width=0.4\textwidth]{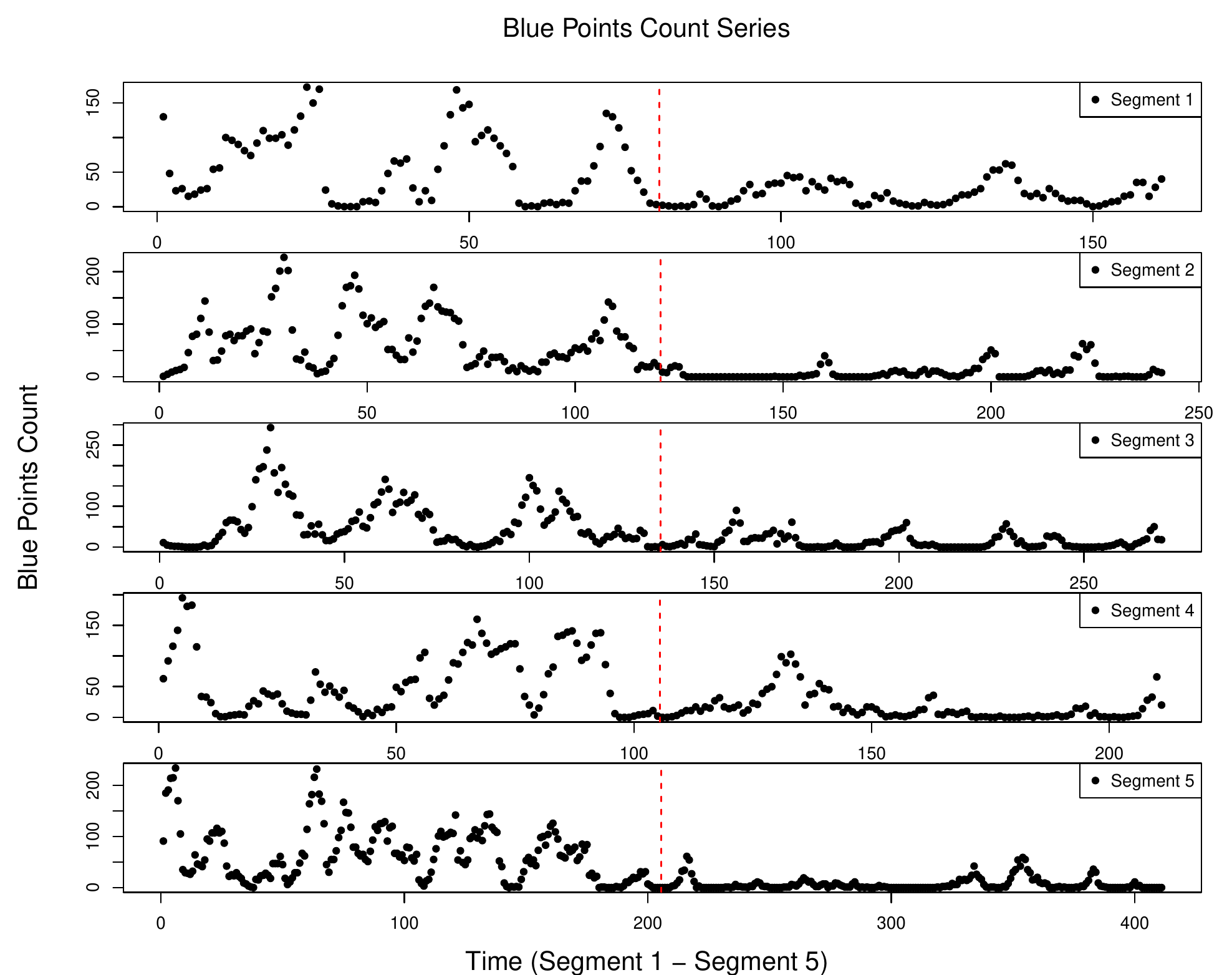}
    \caption{Blue Count Series}
    \label{fig:bluecount}
 \end{figure}

As revealed in the heatmaps, more blue dots are around the first half of the inter-ictal segments but gradually disappear in the later half of segments. One way to summarize this pattern is to count blue dots along the temporal axis. Figure~\ref{fig:bluecount} shows the blue points count series for five segments. All five count series share a common feature, they all have recurrent cycles in the first half segment but their cyclic pattern seems to disappear afterward. The clear pattern difference between the first and second half of inter-ictal segments manifests the neuron system has changed. At the meantime, in each inter-ictal period, the disappearance of recurrent cycles of blue points count strongly indicates that there is a systemic change-point involved, and the change-point is located in the middle of the period. If we could detect the systemic change-point, then we would be able to predict next epileptic seizure.

\begin{figure}[htp]
  \centering
  \subfigure[]{\includegraphics[scale=0.4]{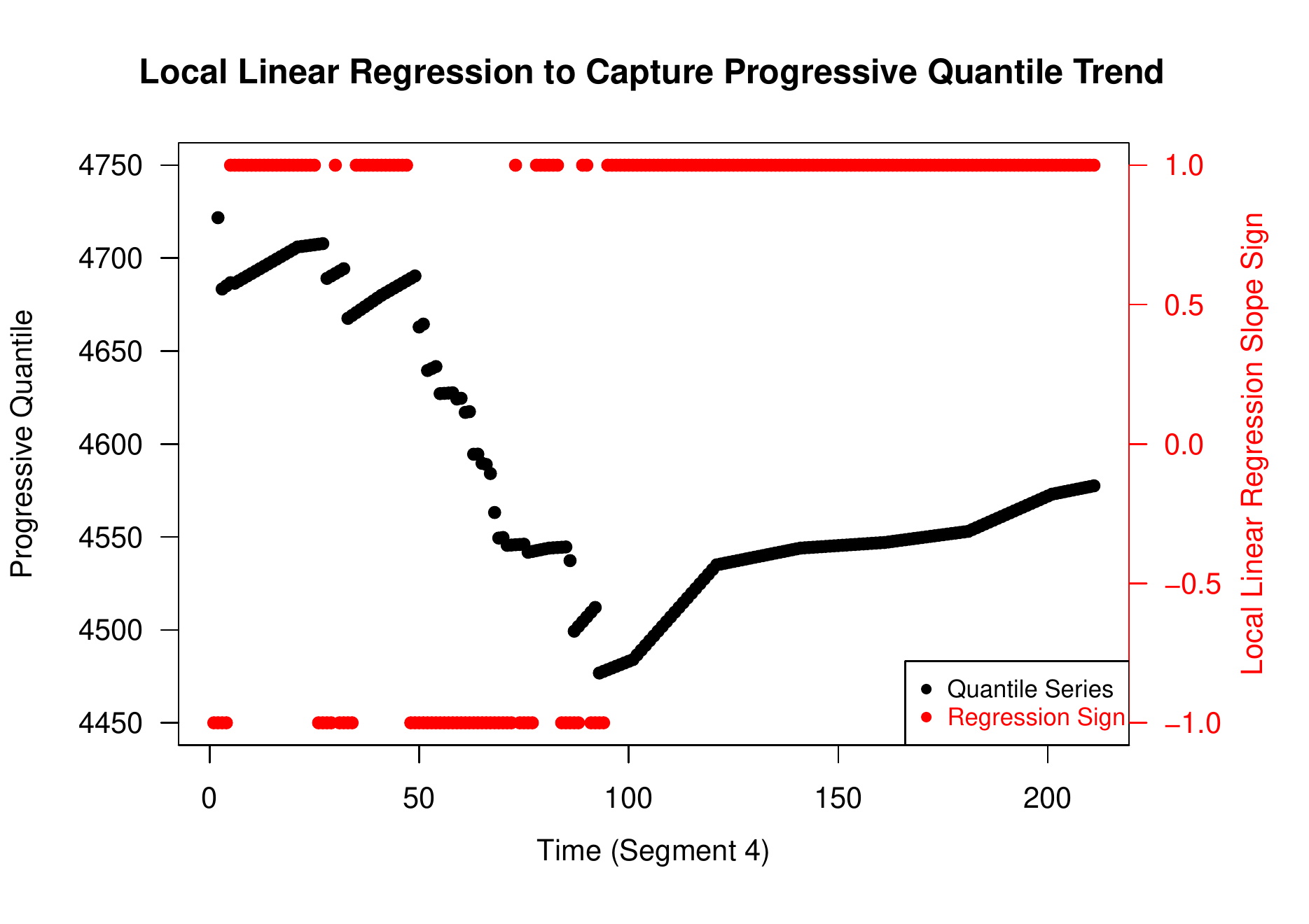}} \quad
  \subfigure[]{\includegraphics[scale=0.28]{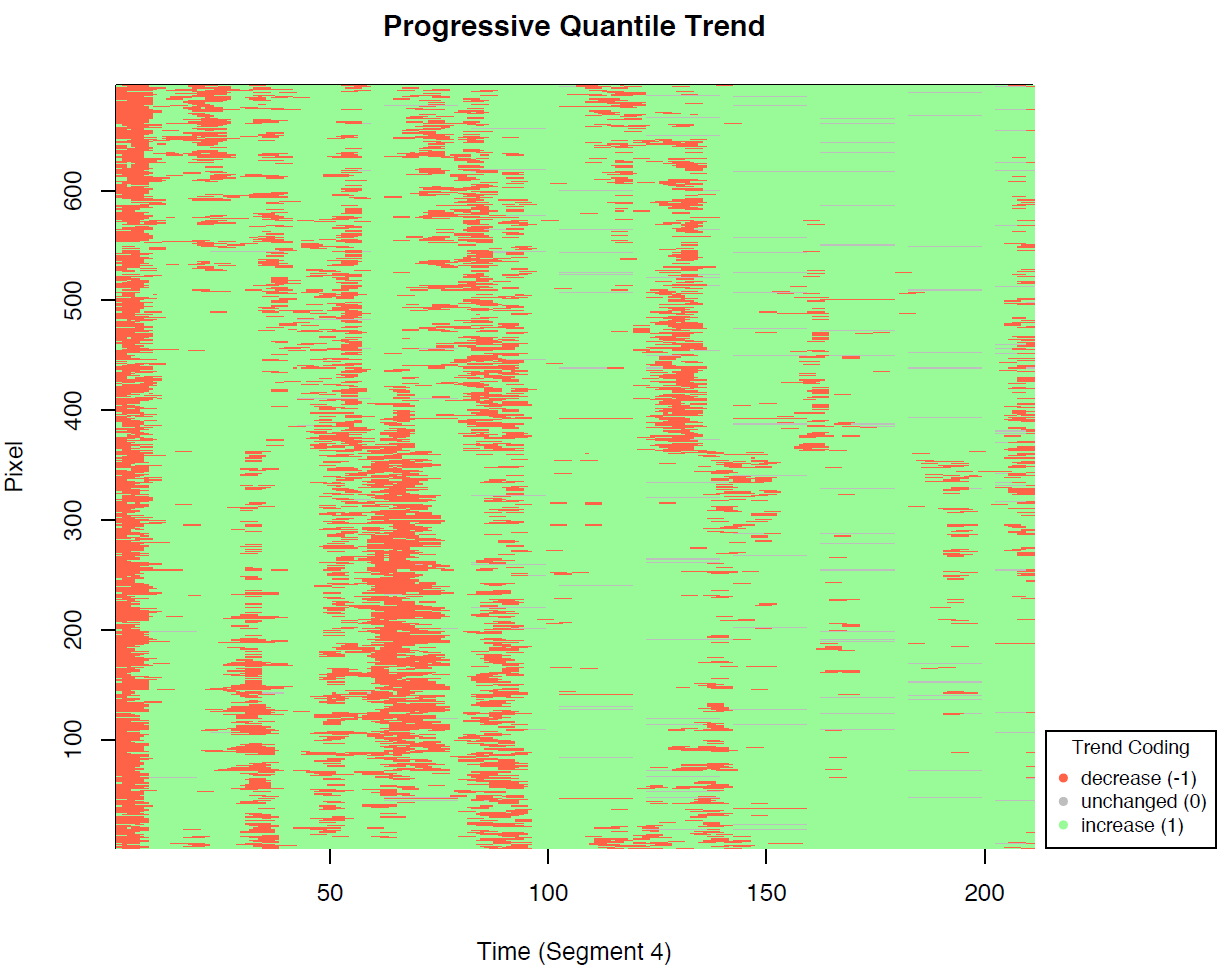}}
  \caption{Progressive Quantile Series Trend}
    \label{fig:sign}
\end{figure}

\begin{figure}[!tbp]
  \centering
  \subfigure[]{\includegraphics[scale=0.35]{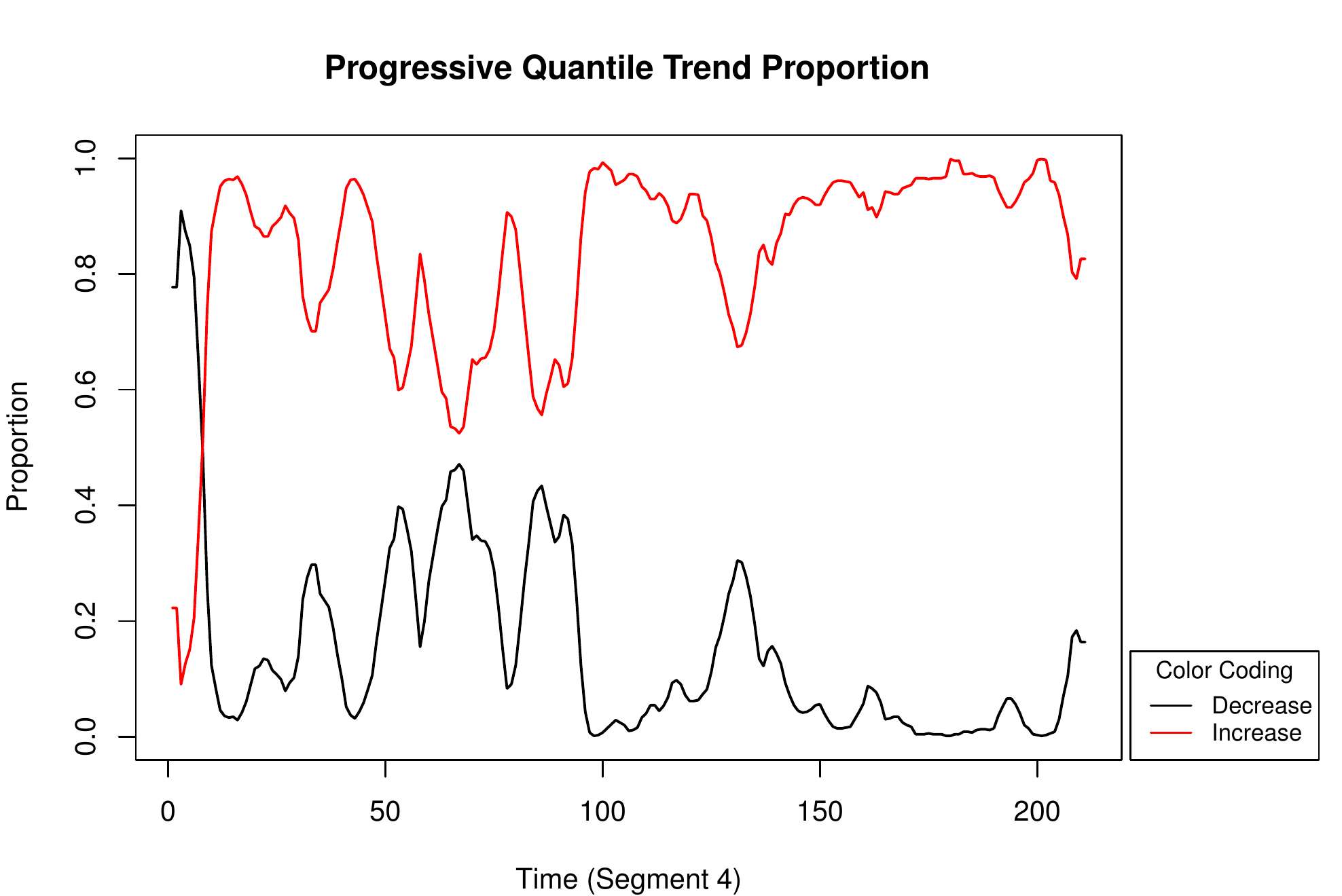}} \quad
  \subfigure[]{\includegraphics[scale=0.33]{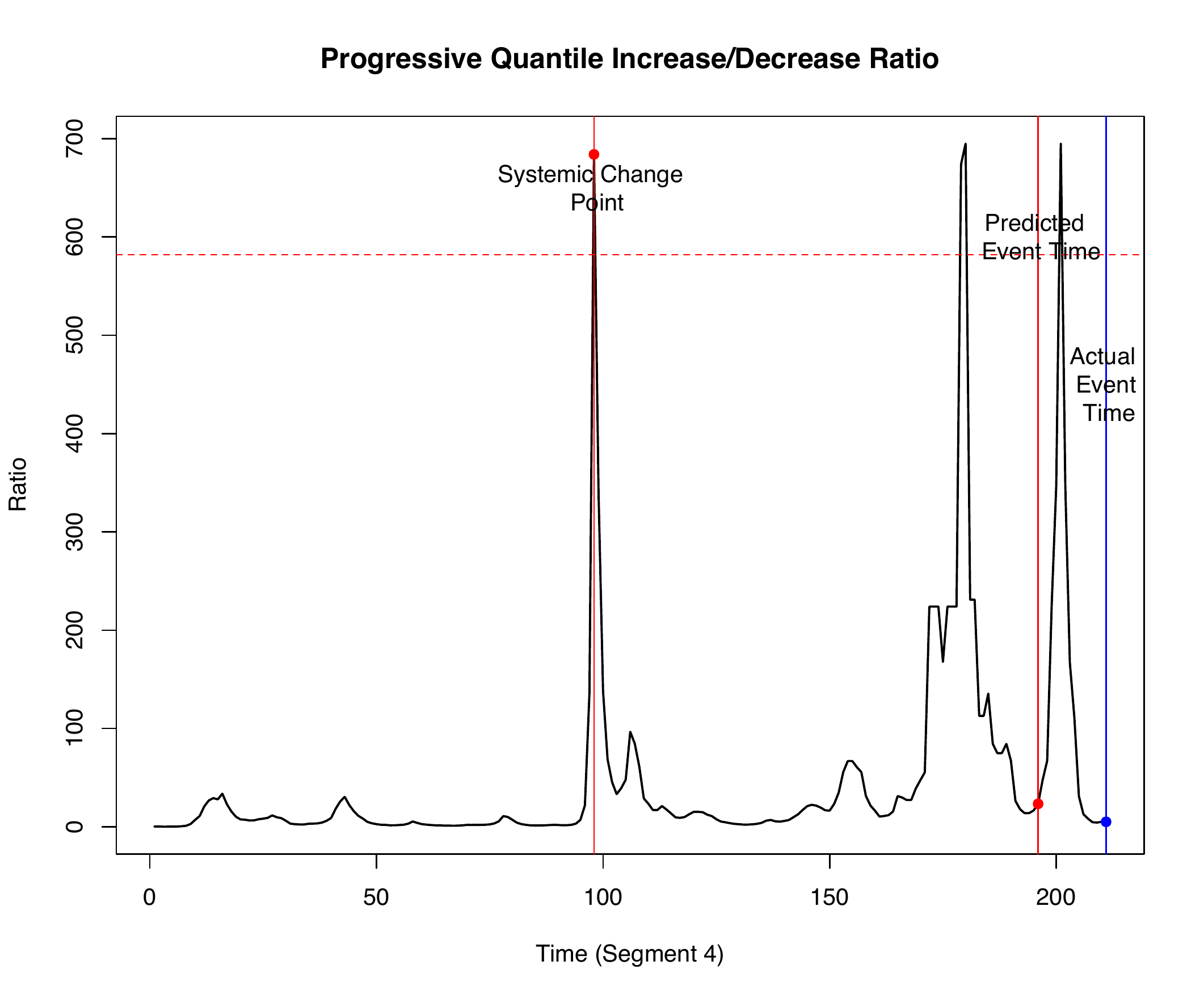}} \quad
   \caption{Prediction Algorithm}
    \label{fig:prediction}
\end{figure}

\subsection{Epileptic Seizure Prediction}

\begin{algorithm}
\SetKwInOut{Input}{input}
\SetKwInOut{Output}{output}
\SetKwInOut{Parameter}{parameter}
\SetAlgoLined
\Input{A calcium fluorescence intensity (CFI) matrix $M$ with 696 rows from picked pixels and T columns from a inter-ictal segment}
\Output{System change point and predicted event time}
$m_i$ = $i^{th}$ row in $M$ = CFI series with length T for $i^{th}$ pixel, $i = 1, \ldots\, 696$\\

\ForEach{$m_i$}{
\ForEach{$t = 1,2, \ldots\, T$}{
    \begin{enumerate}
    \item  Based on $m_i(0), \ldots\, m_i(t)$, \\compute $5\%$ quantile ${D_i^{-1}(0.05|t)}$
    \item  Fit linear regression on $\{{D_i^{-1}(0.05|t-2)}, {D_i^{-1}(0.05|t-1)},
    {D_i^{-1}(0.05|t)}$, ${D_i^{-1}(0.05|t+1)}, {D_i^{-1}(0.05|t+2)}\}$,
    and take the sign of regression slope as $S_i(t)$
    \end{enumerate}
    }
}

\ForEach{Column $S(t)$ in matrix $S$, $t = 1,2, \ldots\, T$}{
    \begin{enumerate}
    \item Count number of 1 and number of -1
    \item Compute the ratio $r(t) = \frac{number of 1}{number of -1}$
    \end{enumerate}
}
Plot $r(t)$ versus time $t$, and record the first time point $r(t)$ is greater than $582$ as the system change point $\hat{T}_{SCP}$.
Predicted epileptic event time $\hat{T}$ is $2*\hat{T}_{SCP}$.
\label{algocompress1}
\caption{Epileptic Event Prediction Algorithm}
\end{algorithm}

To predict a change-point for a time series, we apply progressive type of statistics, such as ${D_i^{-1}(0.05|t)}$, the progressive quantile series. In section ~\ref{quantile}, we show the pattern for ${D_i^{-1}(0.05|t)}$ pixel-wise. For the majority of pixels, their progressive quantile series would result in an increasing trend after systemic change-point. 
To capture such pattern change of progressive quantile series among $696$ pixels, we implement local linear regression with a moving window to measure increase and decrease trend of quantile series. Specifically, we choose a small window with five time points
$$\{{D_i^{-1}(0.05|t-2)}, {D_i^{-1}(0.05|t-1)}, {D_i^{-1}(0.05|t)}, $$
$${D_i^{-1}(0.05|t+1)}, {D_i^{-1}(0.05|t+2)}\},$$
and fit linear regression model. We then label the increase and decrease trend of the quantile series ${D_i^{-1}(0.05|t)}$ as $1$ and $-1$, the sign of regression slope, respectively. Figure \ref{fig:sign} panel (a) shows the progressive quantile series for one pixel and the sign of slope in regression model to indicate the increase or decrease trend.

Hence, for each $696$ pixel, its progressive quantile series ${D_i^{-1}(0.05|t)}$ is transformed into a trend series $\{-1,0, 1\}$. Then we aggregate $696$ trend series in a heatmap, as shown in Figure~\ref{fig:sign} panel (b). The heatmap again shows that the majority of decrease trend appears in the beginning of inter-ictal segment and gradually disappears.
To summarize this pattern change, we calculate the increase(decrease) proportion in time domain, which is the number of pixels that have positive(negative) regression slope over $696$ pixels. The panel (a) of Figure \ref{fig:prediction} shows that the overall decrease trend could form cyclic pattern during the first half inter-ictal segment and the cyclic pattern gradually disappears approaching next epileptic event. We could observe similar results for increase trend. The disappearing cyclic pattern for increase(decrease) trend manifests that the system has changed. Therefore, one reasonable method for detecting the systemic change-points is to trace the time point when the increase or decrease trend loses its cyclic pattern.

To accurately detect the systemic change-point, we compute the increase to decrease ratio, as shown in panel (b) of Figure \ref{fig:prediction}. The increase to decrease ratio indeed enlarges the contrasting difference of cyclic pattern change. Figure \ref{fig:prediction} panel (b) shows that there is a dramatical spike in the middle of inter-ictal period. This spike rightly indicates the systemic change-point. Before the high spike, the ratio has slight fluctuations occasionally. However, after the high spike, the ratio is no longer stable. It strongly fluctuates and forms multiple significant spikes. To successfully locate the high spike, we calculate three standard deviations of the mean of ratio series for five segments, and set the average $582$ as the threshold. Thus, our predicted systemic change-point $\hat{T}_{SCP}$ is the time point when the increase to decrease ratio series first reach $582$.

Once the systemic change-point can be predicted, it is feasible to predict epileptic seizure. Based on blue points count series in Figure \ref{fig:bluecount}, we know that the systemic change-point $T_{SCP}$ is estimated in the middle of the inter-ictal period, which means we could predict epileptic event time $T$ by $2*T_{SCP}$. The increase to decrease ratio gives the estimated systemic change-point $\hat{T}_{SCP}$, then $2*\hat{T}_{SCP}$ is our predicted time for next epileptic event $\hat{T}$. 

\section{Results}\label{Results}

\begin{figure}[!tbp]
  \centering
  \subfigure[Prediction]{\includegraphics[scale=0.33]{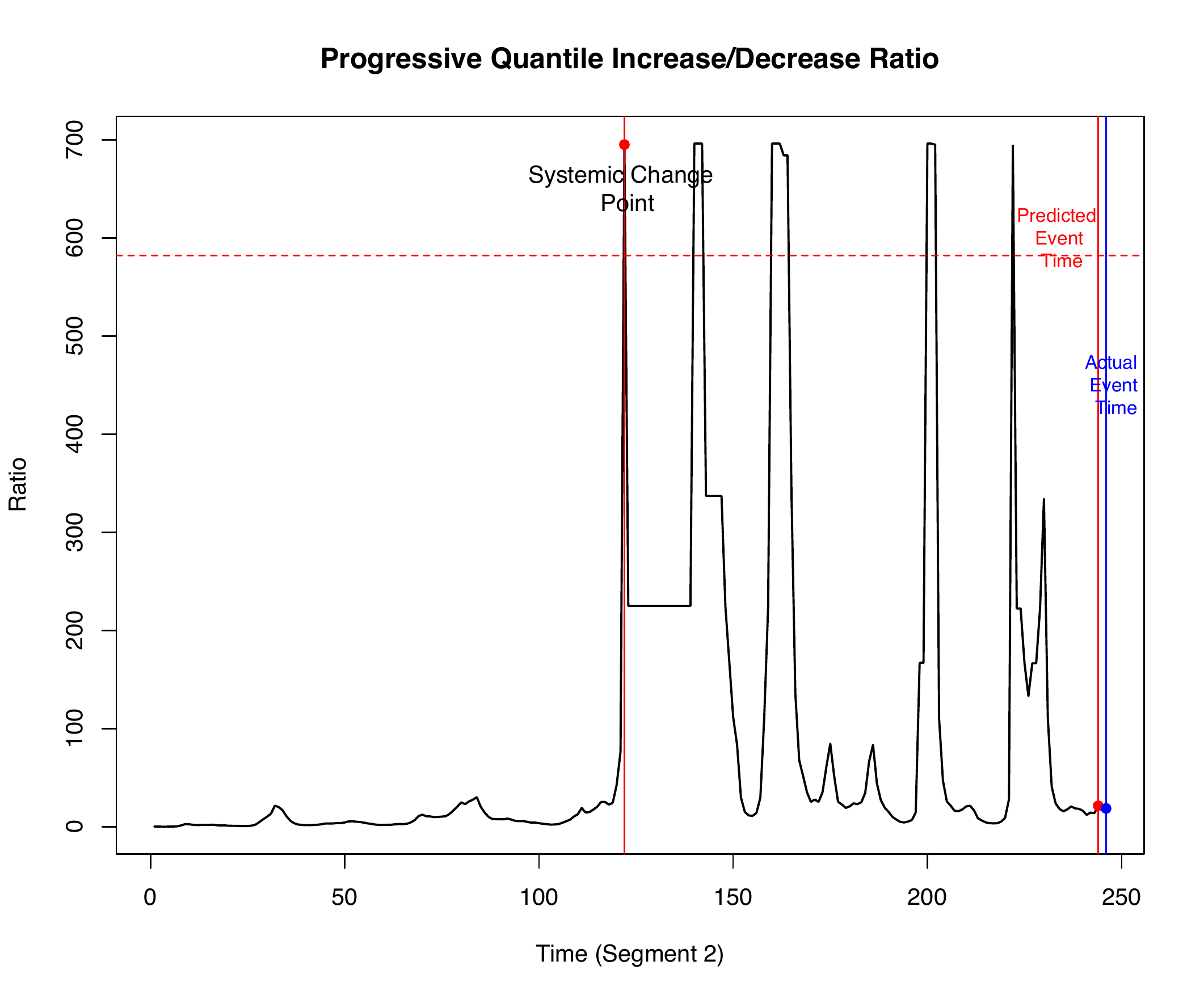}} \quad
  \subfigure[]{\includegraphics[scale=0.26]{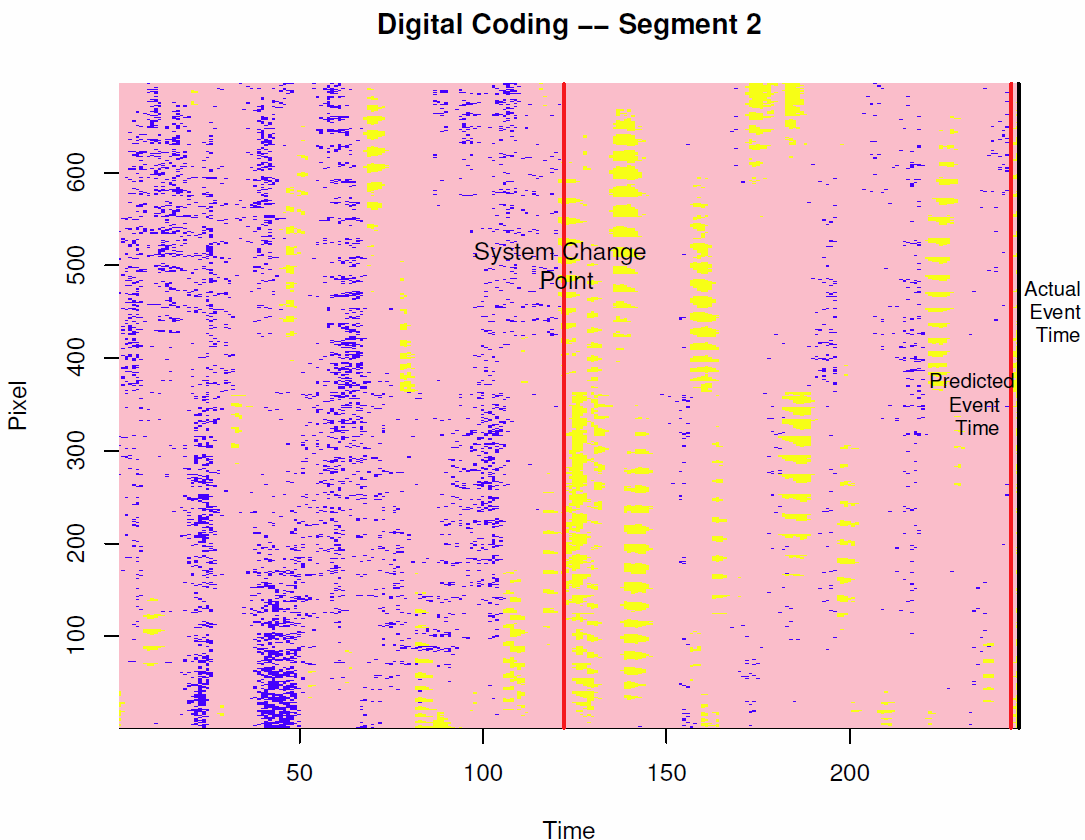}}
   \caption{Prediction (Segment 2)}
    \label{fig:newprediction}
\end{figure}

To evaluate the prediction performance of systemic change-point $T_{SCP}$, one way is to check if the system actually changes after the change-point. To do so, we mark $\hat{T}_{SCP}$ and $\hat{T}$ on the ratio series and digital encoding heatmap for another inter-ictal segment, segment $2$, in Figure \ref{fig:newprediction}. The ratio series in panel (a) has nearly no fluctuations before predicted change-point $\hat{T}_{SCP}$, but forms dramatic spikes afterwards. The patterns of ratio series before and after $\hat{T}_{SCP}$ are clearly different. The heatmap in Figure \ref{fig:newprediction} panel (b) also shows the pattern difference before and after predicted change-point. As we can observe in the heatmap, there are alternate blue and yellow points before $\hat{T}_{SCP}$, but blue points gradually disappear while the intensity of yellow points gradually become higher after $\hat{T}_{SCP}$.

\begin{figure}[!tbp]
  \centering
    \includegraphics[width=0.4\textwidth]{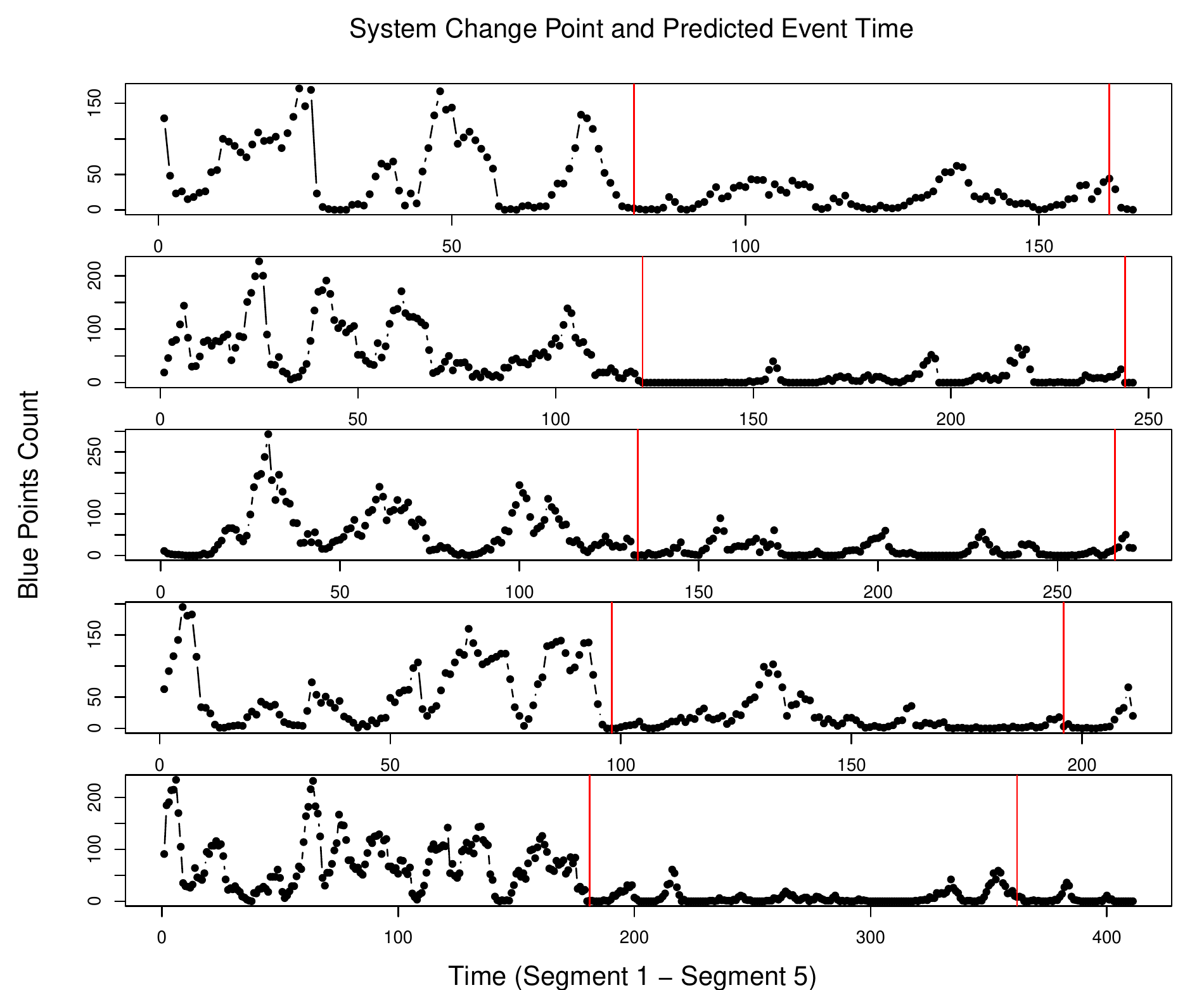}
    \caption{Prediction Result}
    \label{fig:result}
 \end{figure}

Figure \ref{fig:newprediction} also shows the epileptic seizure prediction result, with predicted event time $\hat{T}$ and true event time $T$ marked on ratio series and digital encoding heatmap. In general, our predicted event time is 2.24 seconds (4.82 \% of the segment length) earlier than the actual event time. The prediction results for all inter-ictal segments are summarized in Table \ref{tab:result}. The predicted systemic change-points $\hat{T}_{SCP}$ and predicted event time $\hat{T}$ are marked on the blue points count series in Figure \ref{fig:result} as well.

\begin{table*}
\caption{Prediction Result}\footnotesize
\label{tab:result}
\centering
\begin{tabular}{ccccc}
\hline
                        & $T$          &   $\hat{T}$   &   Absolute Prediction Error  &  Relative Prediction Error  \\
\hline
$Segment 1$  &  $166 (24.78s)$     &     $162 (24.18s) $     &       4 (0.6s)    &     2.41\%     \\
$Segment 2$  &  $246 (36.72)$     &     $244 (36.42)$     &       2 (0.3s)         &    0.81\%     \\
$Segment 3$  &  $271 (40.45)$     &     $266 (39.70)$     &       5 (0.75s)       &    1.85\%     \\
$Segment 4$  &  $211 (31.49)$     &     $196 (29.25)$     &       15 (2.24s)      &   7.11\%      \\
$Segment 5$  &  $411 (61.34s)$     &     $362 (54.03)$     &       49 (7.31s)     &  11.92\%     \\
\hline
Mean Error     &      &           &        $2.24s$            &       $ 4.82\% $            \\
\hline
\end{tabular}
\end{table*}

\section{Conclusion}\label{con}
Techniques of data driven computing and graphic display are devised to coherently represent systemic patterns of epileptic mechanism based on one Zebrafish's brain-wide calcium imaging video. Our data-driven computing via the pixel-specific progressive quantile process successfully extracts authentic information content that is pertinent to the target epileptic mechanism. This aspect is rather distinct to classic modeling methodologies developed for analyzing EEG data.

Such a progressive quantile statistics then motivates a triplet digital encoding scheme to renormalize all selected informative pixels' calcium intensity time series into digital sequences. Within each inter-ictal period, all digital sequences collectively reveal the clear systemic patterns of wax and wane of inhibition and excitation along the temporal axis. Very importantly a systemic change-point is observed evidently. Such a visible change-point during the evolving processes of troughs and peaks of original calcium intensities belonging to all pixels could be rather difficult to visualize due to their volatility in values and in ranges.

Further we discover that the increasing and decreasing trends of progressive quantile series also collectively recover the systemic patterns of triplet-digital encoding sequences within each inter-ictal period. We then devise a systemic change-point estimation, which, as an early warning signal, can capture rather well the arrival time of incoming epileptic events.
Throughout this paper, the graphic displays constantly play a central role of fostering understanding of information contained in the video data. Such understanding not only can help biologists and neuroscientists to discover new insights, but also help computing scientists to be sure that right information contents are extracted.


\section*{Acknowledgment}
We would like to thank Professor Scott Baraban at University of California, San Francisco for providing the zebrafish brain-wide calcium imaging video data.

\bibliographystyle{ACM-Reference-Format}
\bibliography{sample-bibliography}

\end{document}